\documentclass[aps,prd,reprint,superscriptaddress,amsmath,amssymb,nofootinbib,nolongbibliography,floatfix]{revtex4-2}
\usepackage{graphicx}
\usepackage{dcolumn}
\usepackage[utf8]{inputenc}

\usepackage{ulem}
\usepackage{xcolor}
\usepackage[unicode=true,colorlinks,linkcolor=blue,anchorcolor=blue,urlcolor=blue,citecolor=blue,breaklinks=true]{hyperref}
\usepackage{subfigure}

\newcommand{\ii}{\mathrm{i}\,}

\newcommand{\pararrow}{\mathord{\buildrel{\lower3pt\hbox{$\scriptscriptstyle\leftrightarrow$}}\over {\partial}}} 
\newcommand{\pararrowk}[1]{\mathord{\buildrel{\lower3pt\hbox{$\scriptscriptstyle\leftrightarrow$}}\over {\partial}\hspace*{-0.18em}{}^#1}\hspace*{-0.18em} \,} 
\usepackage{bm}

\newcommand{\qfnu}{\affiliation{College of Physics and Engineering, Qufu Normal University, Qufu 273165, China}}

\begin{document}

\title{Light hadronic decays of spin-0 partner of $X(3872)$}

    \author{Hong-Shuo Gao} \qfnu
    \author{Zu-Xin Cai} \qfnu
    \author{Gang Li}\email{gli@qfnu.edu.cn (Corresponding author)} \qfnu
    \author{Shi-Dong Liu}\email{liusd@qfnu.edu.cn (Corresponding author)} \qfnu 

\begin{abstract}

The $X(3872)$ is theoretically predicted to have a spin-0 partner state denoted as $X_0$. Assuming the $X_0$ as a $D\bar{D}$ molecular bound state, we calculate the decay widths of $X_0\to VV$ and $X_0 \to PP$ ($V$ and $P$ stand for light vector and pseudoscalar mesons, respectively) via intermediate charmed meson loops. Three different configurations of the $X_0$, i.e., pure neutral components ($\theta = 0$), isospin singlet ($\theta = \pi/4$) and pure charged components ($\theta = \pi/2$), are investigated. Within a commonly accepted range of the model parameter $\alpha$, the predicted decay widths of $X_0 \to VV$ are on the order of a few hundred $\mathrm{keV}$, while the decay widths of $X_0 \to PP$ can reach several $\mathrm{MeV}$. The $X_0 \to \rho\rho$ and $\pi\pi$  have larger decay rates. The relative width ratios between the channels are nearly model-independent. Moreover, among those channels only with isovector or isoscalar mesons, the relevant ratios are also independent of the phase angle. The predicted ratios are helpful for searching the $X_0$ in the future experiments at BESIII and Belle II.
\\
\\
\textbf{Keywords:} Intermediate meson loops, Light hadronic decays, Molecular states
\end{abstract}

\date{\today}

\maketitle


\section{Introduction} \label{sec:introduction}

In 2003, the Belle Collaboration observed the $X(3872)$ resonance in the $\pi^+\pi^-J/\psi$ invariant mass spectrum from the $B \to K \pi^+ \pi^- J/\psi$ process~\cite{Belle:2003nnu}, marking a pivotal breakthrough in exotic hadron spectroscopy. The LHCb Collaboration determined its quantum number $J^{PC}=1^{++}$~\cite{LHCb:2013kgk,LHCb:2015jfc}. At present, the world average mass of the $X(3872)$ is $(3871.64\pm 0.06)~\mathrm{MeV}$, with an extremely narrow total width of $(1.19\pm 0.21)~\mathrm{MeV}$~\cite{ParticleDataGroup:2024cfk}. The mass of $X(3872)$ is very close to the threshold of $D^0 \bar{D}^{*0}$ threshold ($m_{D^0}+m_{D^{*0}} = 3871.69~\mathrm{MeV}$), and its dominant decay channels is $D \bar{D}^{*}$ final states. Therefore, the $X(3872)$ can be naturally interpreted as the $D \bar{D}^{*}$ molecular state~\cite{Close:2003sg,Swanson:2003tb,Wong:2003xk,Suzuki:2005ha,Ebert:2005nc,Swanson:2004pp,Fleming:2007rp,Bignamini:2009sk,Nieves:2012tt,AlFiky:2005jd,Hanhart:2007yq,Gamermann:2009fv,Braaten:2007dw,Hidalgo-Duque:2012rqv,Dong:2008gb,Braaten:2006sy,Wang:2013daa,Kang:2016jxw,Fleming:2008yn,Guo:2014hqa,Ding:2009vj,Lee:2009hy,Guo:2014taa,Braaten:2003he,Guo:2017jvc}. Meanwhile, there also exist other explanations, such as tetraquark state~\cite{Chen:2019eeq,Shi:2021jyr,Wang:2013vex,Maiani:2014aja,Wang:2013vex,Barnea:2006sd,Maiani:2004vq,Wang:2023sii}, hybrid charmonium
state~\cite{Close:2003mb,Li:2004sta}, charmonium-molecule mixing state~\cite{Chen:2013pya,Takizawa:2012hy}, and conventional
charmonium state~\cite{Barnes:2003vb,Quigg:2004vf,Achasov:2024ezv,Achasov:2024anu,Eichten:2004uh}.

Heavy Quark Spin Symmetry (HQSS) provides a crucial theoretical framework for understanding the properties and interactions of hadrons containing heavy quarks~\cite{Yan:1992gz,Neubert:1993mb,Isgur:1989vq,Fleming:2008yn}. According to the effective field theory (EFT) and HQSS, the $X(3872)$ should have its HQSS partners~\cite{Nieves:2012tt}. If the $X(3872)$ is $D\bar{D}^*$ molecular state with $J^{PC} = 1^{++}$, there would exist three degenerate partner states with the quantum numbers $1^{+-}$, $0^{++}$ and $2^{++}$ using an EFT approach with contact interactions in the $D \bar{D}$, $D \bar{D}^{*}$, and $D^* \bar{D}^{*}$ channels in the heavy quark limit~\cite{Baru:2017pvh}. 

Extensive investigations have been conducted on the characteristics and structure of the $X_2$, the isoscalar $2^{++}$ $D^* \bar{D}^{*}$ partner of the $X(3872)$~\cite{Albaladejo:2015dsa,Shi:2023mer,Liu:2024ogo,Ganbold:2024buy,Shi:2023ntq,Montana:2022inz,Liu:2019stu,Guo:2013sya}. The mass of $X_2$ is theoretically predicted to be around $4012~\mathrm{MeV}$ with a width of approximately the same magnitude as that of the $X(3872)$~\cite{Hidalgo-Duque:2012rqv,Guo:2014ura,Baru:2016iwj,Baru:2017fgv,Baru:2024ptl,Wang:2020dgr,Shi:2024llv}. In 2022, Belle collaboration observed a new structure with a mass of $(4014.3\pm 4.0 \pm 1.5)~\mathrm{MeV}$ and a width of $(4\pm 11\pm 6)~\mathrm{MeV}$ in the invariant mass distribution of the $\gamma \psi(2S)$~\cite{Belle:2021nuv}. Given that the mass and width of the newly-discovered structure are in accordance with the predicted mass and width of the $X_2$, it is considered to be a good candidate for the $X_2$. 
In Ref.~\cite{Zheng:2024eia}, the hidden charmed decays of the $X_2 \to J/\psi V$ and $X_2 \to \eta_c P$ via charmed meson loops were discussed, where $V=\rho^0$, $\omega$, and $P=\pi^0$, $\eta$, and $\eta^\prime$. The results indicate that the decay widths are significantly influenced by the $X_2$ mass. In Ref.~\cite{Cai:2024glz}, the charmless decays of the $X_2 \to VV$ and $X_2 \to PP$, where $V$ represents light vector and $P$ stands for pseudoscalar were investigated. In the cases where the $X_2$ is a purely neutral $D^{*0} \bar{D}^{*0}$ or a purely charged $D^{*+} D^{*-}$ bound state, the theoretically predicted partial decay widths for the $X_2 \to VV$ and $X_2 \to PP$ are at order of several tens of $\mathrm{keV}$.

The spin-0 partner of $X(3872)$, denoted as $X_0$ ($J^{PC}=0^{++}$), is initially predicted using a coupled channel unitary approach~\cite{Gamermann:2006nm} involving many $PP$ channels. The only relevant Experimental studies are those by the Belle and BaBar Collaborations, which analyzed the $e^+ e^- \to J/\psi D \bar{D}$ and  $e^+ e^- \to  D \bar{D}$ reactions~\cite{Belle:2007woe,Belle:2017egg,BaBar:2010jfn}. 
However, numerous theoretical studies have been carried out in different processes~\cite{Gamermann:2007mu,Wang:2019evy,Nieves:2012tt,Hidalgo-Duque:2012rqv,Ding:2020dio,Xiao:2012iq,Brandao:2023vyg,Dai:2020yfu,Prelovsek:2020eiw,Wang:2020elp,Deineka:2021aeu,Dai:2015bcc}. For example, in the $D \bar{D}$ mass distribution of $e^+ e^- \to J/\psi D \bar{D}$ reaction~\cite{Gamermann:2007mu,Wang:2019evy}; in the analysis of meson-meson interactions within the heavy meson effective theory~\cite{Nieves:2012tt,Hidalgo-Duque:2012rqv,Ding:2020dio}; in the $D^0 \bar{D}^0$ mass distribution of the $\psi (3770) \rightarrow \gamma D^0 {\bar{D}}^0$ decay~\cite{Dai:2020yfu}. In~\cite{Wang:2020elp,Deineka:2021aeu},  the authors described the $D{\bar D}$ mass  distribution of $\gamma \gamma \to D{\bar D}$ and demonstrated the existence of a bound state near the $D{\bar D}$ mass threshold. Dai $et$ $al$. predicted the $D \bar{D}$ mass distribution in $B^-\to D{\bar D} K^-$ process by considering  the state $X(3720)$ mainly coupled to $D \bar{D}$~\cite{Dai:2015bcc}. The width of the $D\bar{D}$ scalar meson $X(3700)$ to the $\eta_c\eta$ channel is $0.85~\mathrm{MeV}$~\cite{Xiao:2012iq}. According to the heavy quark spin symmetry and coupled-channel dynamics, the $0^{++}$ hadronic molecule can not only be a $D\bar{D}$ molecule bound state, but also be regarded as a $D^*\bar{D}^*$ state, or even a mixture of $D\bar{D}$ and $D^*\bar{D}^*$~\cite{Nieves:2012tt,Baru:2024ptl}. However, this work considers only the case where $0^{++}$ molecule is composed entirely of $D\bar{D}$. The global analysis of measurements from the Belle, BABAR, BESIII, and LHCb experiments implies that the fraction of unobserved decay modes of the $X(3872)$ is up to $31.9\%$, which indicates that there is still a lot of room for searching for new decay modes of the $X(3872)$ \cite{Li:2019kpj}, for example the light hadron decays \cite{BESIII:2023xta,Wang:2022qxe}. The $X_0$, as the spin-0 partner of $X(3872)$, might have some common decay properties with the $X(3872)$. In continuation of our previous work devoted to the light hadron decays of the exotic states \cite{Wang:2022qxe,Wu:2016ypc,Wu:2022hck,Cai:2024glz}, we shall investigate the light hadronic decays $X_0 \to VV$ and $X_0 \to PP$ via the intermediate charmed meson loops. The calculations were performed under the $D\bar{D}$ molecule ansatz of the $X_0$ and using an effective Lagrangian approach.

The rest of the paper is organized as follows. In Sec.~\ref{sec:formalism}, we provide the relevant effective Lagrangians and expressions for the decay amplitudes. Then in Sec.~\ref{sec:results}, the numerical results and discussions are presented, and a brief summary is given in Sec.~\ref{sec:summary}.

\section{Theoretical Framework} \label{sec:formalism}

\subsection{Effective interaction Lagrangians} \label{subsec:2.1}

We assume that $X_0$ is an $S$-wave molecular state with the quantum numbers $J^{PC}=0^{++}$, which can be given by the superposition of the hadronic configurations of $D^0\bar{D}^0$ and $D^+D^-$
\begin{equation}\label{eq:WX0}
    |X_0 \rangle = \cos{\theta}|D^0{\bar D}^0 \rangle + \sin{\theta}|D^+ D^- \rangle,
\end{equation}
where $\theta$ represents a phase angle that defines the proportion of the neutral and charged components in the system. The interaction between the $X_0$ state and the $D\bar{D}$ meson pair can be described by the Lagrangian
\begin{equation}\label{eq:LX0}
    \mathcal{L}_{X_0} = X_0\left(\chi_\mathrm{nr}^0  D^{0 \dagger}\bar{D}^{0 \dagger}\cos{\theta} + \chi_{\mathrm{nr}}^\mathrm{c} D^{+ \dagger}D^{- \dagger}\sin{\theta}\right),
\end{equation}
where $\chi_{\mathrm{nr}}^0$ and $\chi_{\mathrm{nr}}^c$ represent the coupling constants of the $X_0$ with the neutral and charged meson pairs, respectively. Their values are related to the $X_0$ binding energy and estimated by \cite{Weinberg:1965zz,Baru:2003qq,Guo:2013zbw,Shi:2023mer}
\begin{equation}\label{eq:chi}     
\chi_{\mathrm{nr}}^2 = \frac{16\pi}{\mu}\sqrt{\frac{2E_\mathrm{B}}{\mu}}, 
\end{equation}
where $E_\mathrm{B}=m_1+m_2-M$ and $\mu=m_1m_2/(m_1+m_2)$. 
Considering that the $X_0$ is a pure $D\bar{D}$ bound state and the mass of the $X_0$ is $3721~\mathrm{MeV}$~\cite{Wei:2022jgc}, we obtain  
\begin{eqnarray}
    \left|\chi_{\mathrm{nr}}^0\right| &=& 2.71~\mathrm{GeV^{-1/2}},\label{eq:chi0} \\
    \left|\chi_{\mathrm{nr}}^c\right| &=& 3.26~\mathrm{GeV^{-1/2}}. \label{eq:chic}
\end{eqnarray}
Such a distinction is due to the mass difference between the neutral and charged mesons, which would enhance the isospin-breaking effect.

\begin{figure}
	\includegraphics[width=0.94\linewidth]{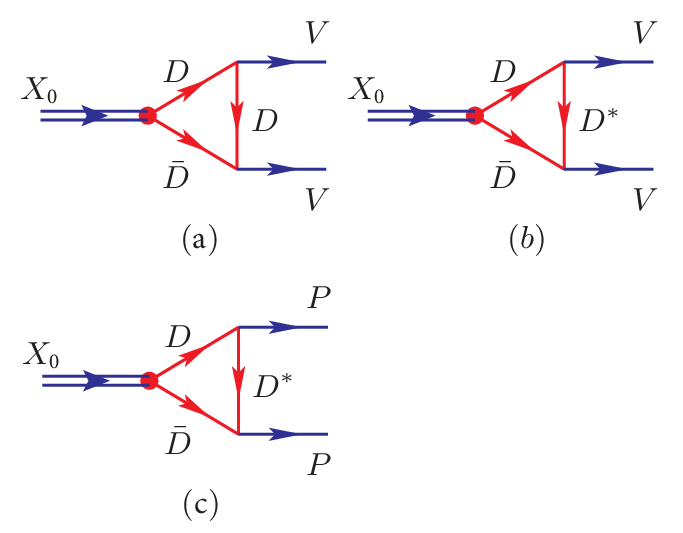}
    \caption{Feynman diagrams for the processes $X_0 \to VV$[(a) and (b)] and $X_0 \to PP$ (c) via intermediate charmed meson loops. The charge conjugated loops are not shown here, but included in the calculations.}
    \label{fig:Feynmandiagrams}
\end{figure}

Furthermore, we assume that the $X_0$ decays to $VV$ and $PP$ ($V$ and $P$ represent light vector and pseudoscalar mesons, respectively) through intermediate charmed meson loops, as shown in Fig.~\ref{fig:Feynmandiagrams}. First, the $X_0$ transforms into $D\bar{D}$ meson pair. Subsequently, the $D$ and $\bar{D}$ evolve into the final states $VV$ or $PP$ via exchanging a charmed meson. When the final states are $VV$, the exchanged charmed meson can be either $D$ or $D^*$, whereas for the $PP$ final state, the exchanged meson is exclusively be $D^*$.

Under the heavy quark limit and chiral symmetry, the interactions of the light vector and pseudoscalar mesons with the charmed mesons are described by the following Lagrangian~\cite{Yan:1992gz,Burdman:1992gh,Wise:1992hn,Casalbuoni:1996pg,Cheng:2004ru}
\begin{align}\label{eq:LDDVP}
\mathcal{L} =& -\ii g_{DDV} D_i^\dagger \pararrowk{\mu}D^{j}(V_\mu^\dagger)_j^i\nonumber\\
&-2 f_{D^*DV}\epsilon_{\mu\nu\alpha\beta} (\partial^\mu V^{\nu\dagger})_j^i(D_i^\dagger \pararrowk{\alpha}D^{*\beta j}- D_i^{*\beta\dagger}\pararrowk{\alpha}D^{j}) \nonumber \\ & +i g_{D^*D^*V}D_i^{*\nu\dagger}\pararrowk{\mu}D_\nu^{*j}(V_\mu^\dagger)_j^i \nonumber\\
&+  4\ii f_{D^*D^*V} D_{i\mu}^{*\dagger}(\partial^\mu V^{\nu\dagger} - \partial^\nu V^{\mu\dagger})_j^i D_\nu^{*j}\nonumber \\
&-\ii g_{D^*DP}\big(D^{i \dagger}\partial^{\mu} P_{ij}^\dagger D_\mu^{*j} - D_\mu^{*i\dagger}\partial^\mu P_{ij}^\dagger D^j\big) \nonumber\\
&+ \frac{1}{2} g_{D^*D^*P}\epsilon_{\mu\nu\alpha\beta} D_i^{*\mu\dagger}\partial^\nu P^{ij\dagger}\pararrowk{\alpha} D_j^{*\beta},
\end{align}
where $D^{(*)} = (D^{(*)0}, D^{(*)+}, D_s^{(*)+})$ and $D^{(*)\dagger} = (\bar{D}^{(*)0}, D^{(*)-}, D_s^{(*)-})$. The $V$ and $P$ are $3\times3$ matrices representing the light vector and pseudoscalar mesons respectively, and their specific forms are as follows:
\begin{subequations}\label{eq:vmatrix}
\begin{align}
    V &= 
    \begin{pmatrix}
    \frac{\rho^0}{\sqrt{2}}+\frac{\omega}{\sqrt{2}}&\rho^+&K^{*+}\\
    \rho^-&-\frac{\rho^0}{\sqrt{2}}+\frac{\omega}{\sqrt{2}}&K^{*0}\\
    K^{*-}&\bar{K}^{*0}&\phi\label{eq:V0}
    \end{pmatrix},\\
	P &= \begin{pmatrix}
		\frac{\pi^0}{\sqrt{2}} + \frac{\delta \eta + \gamma \eta'}{\sqrt{2}} & \pi^+ & K^+\\
		\pi^- & -\frac{\pi^0}{\sqrt{2}} + \frac{\delta \eta + \gamma \eta'}{\sqrt{2}} & K^0\\
		K^- & \bar{K}^0 & - \gamma \eta + \delta \eta' \label{eq:P0}
	\end{pmatrix}.
\end{align}
\end{subequations}
Here the physical states $\eta$ and $\eta'$ are superposition of $n\bar{n}=(u\bar{u}+d\bar{d})/\sqrt{2}$ and $s\bar{s}$ flavor eigenstates. Their wave functions are given by
\begin{equation}\label{eq:etaetaprime}
	\left( 
	\begin{array}{c}
		\left| \eta\right\rangle  \\
		\left| \eta'\right\rangle 
		\end{array}
	\right) 
	=\left( 
	\begin{array}{cc}
		\delta& -\gamma  \\
		\gamma & \delta
		\end{array}
		\right)
		\left( 
		\begin{array}{c}
		\left| n\bar{n}\right\rangle \\
		\left| s\bar{s}\right\rangle  
		\end{array}
		\right),
\end{equation}
where $\delta=\cos(\theta_\mathrm{P}+\arctan\sqrt{2})$ and $\gamma=\sin(\theta_\mathrm{P}+\arctan\sqrt{2})$ with the $\eta$-$\eta'$ mixing angle $\theta_\mathrm{P}$ varying from $-24.6^\circ$ to $-11.5^\circ$ \cite{ParticleDataGroup:2024cfk}.

The coupling constants of the charmed mesons to the light vector and pseudoscalar mesons are determined as follows~\cite{Casalbuoni:1996pg}:
\begin{subequations}\label{eq:gddvs}
\begin{align}
    g_{DDV} &=  \frac{\beta g_V}{\sqrt{2}},\label{eq:gddvgdsdsv}\\
    f_{D^*DV}&=  \frac{\lambda g_V}{\sqrt{2}},\label{eq:fdsdvfdsdsv}\\
    g_{D^*DP}&= \frac{2g}{f_{\pi}} \sqrt{m_Dm_{D^*}}.\label{eq:gddp}
\end{align}
\end{subequations}
Here $g_V = m_\rho / f_\pi$ with the pion decay constant $f_\pi = 132~\mathrm{MeV}$~\cite{Casalbuoni:1996pg} and the $\rho$ meson mass $m_{\rho}=775.26~\mathrm{MeV}$~\cite{ParticleDataGroup:2024cfk}. Following Ref.~\cite{Isola:2003fh}, $\beta=0.9$, $\lambda=0.56~\mathrm{GeV^{-1}}$, and $g = 0.59$.

\subsection{Transition amplitudes of $X_0 \to V V$ and $X_0 \to P P$} \label{subsec:2.3}

According to the effective Lagrangians above, the amplitudes $\mathcal{M}_V$ for the decay $X_0 \to VV$ through the charmed meson loops in Figs.~\ref{fig:Feynmandiagrams}(a) and (b) are given as
\begin{align}
		\mathcal{M}_V^a = & \int \frac{\mathrm{d}^4q}{(2\pi)^4} [\chi_{\mathrm{nr}}^{0,c }\sqrt{m_{X_0}}m_D \varepsilon^{*\mu}(V_1)\varepsilon^{*\nu}(V_2)]\nonumber\\
  &\times[g_{DDV}(p_1+q)_\mu][g_{DDV}(q-p_2)_\nu]\nonumber\\	
         &\times S(p_1,m_D) S(p_2,m_D) S(q,m_D)\mathcal{F}(q^2),\label{eq:a}\\ 
		\mathcal{M}_V^b  = & \int \frac{\mathrm{d}^4q}{(2\pi)^4} [\chi_{\mathrm{nr}}^{0,c }\sqrt{m_{X_0}}m_D \varepsilon^{*\mu}(V_1)\varepsilon^{*\nu}(V_2)] \nonumber\\
  &\times[2f_{D^*DV}\epsilon_{\delta\mu\omega\xi}p_3^\delta(p_1+q)^\omega]\nonumber\\
  &\times[2f_{D^*DV}\epsilon_{\lambda\nu\gamma\eta}p_4^\lambda(q-p_2)^\gamma]\nonumber\\
        &\times S(p_1,m_D)S(p_2,m_D) S^{\xi\eta}(q,m_{D^*})\mathcal{F}(q^2).\label{eq:b}
\end{align}
The amplitude of the decay $X_0 \to PP$ shown in Fig.~\ref{fig:Feynmandiagrams}(c) is expressed as
\begin{align}
		\mathcal{M}_P^c  = & \int \frac{\mathrm{d}^4q}{(2\pi)^4} [\chi_{\mathrm{nr}}^{0,c }\sqrt{m_{X_0}}m_D ][g_{D^*DP} p_{3\xi}] [-g_{D^*DP} p_{4\eta}]\nonumber\\
  &\times S(p_1,m_D) S(p_2,m_D)S^{\xi\eta}(q,m_{D^*})\mathcal{F}(q^2).\label{eq:c}
\end{align}
In Eqs.~\eqref{eq:a}--\eqref{eq:c} the mass factor $\sqrt{m_{X_0}}m_D$ is responsible for the nonrelativistic normalization of the heavy fields at the $X_0D\bar{D}$ interaction vertex. The symbols $\varepsilon^{*\mu}(V_1)$ and $\varepsilon^{*\nu}(V_2) $ are the polarization vectors of the final states $V_1$ and $V_2$, respectively. The $S(q,m_D)$ represents the propagator for the scalar $D$, while $S^{\mu\nu}(q,m_{D^*})$ stands for the propagator of the vector $D^*$, in the following forms: 
\begin{eqnarray}
		S(q,m_D) &=& \frac{1}{q^2-m_D^2+\ii \epsilon},\\
		S^{\mu\nu}(q,m_{D^*}) &=& \frac{-g^{\mu\nu} + q^\mu q^\nu/m_{D^*}^2}{q^2-m^2_{D^*}+\ii \epsilon}.
\end{eqnarray}

To model the off-shell behavior of exchanged mesons and the internal structure of involved mesons, the form factor $\mathcal{F}(q^2)$ is introduced in Eqs.(10)-(12) [88, 90-93]. The mesons $D$ and $\bar{D}$ interacting with $X_0$ can be considered as on-shell, since the mass of the $X_0$ is close to the $D\bar{D}$ threshold. However, the exchanged meson in the triangle loop is off-shell. For the light hadron decay processes studied in this work, we adopt a tripole form factor
\begin{equation}
    \mathcal{F}(q^2) = \left(\frac{m^2-\Lambda^2}{q^2-\Lambda^2}\right)^3,
    \label{Eq:formfactor}
\end{equation}
where $m$ and $q$ are the mass and momentum of the exchanged meson, respectively. The cutoff $\Lambda$ is reparameterized as $\Lambda = m + \alpha \Lambda_{\mathrm{QCD}}$ with $\Lambda_{\mathrm{QCD}} = 0.22~\mathrm{GeV}$ [88]. In our test calculations, the partial decay widths obtained using monopole or dipole form factors are significantly large, even exceeding the full width of the $X_0$, which is $10~\mathrm{MeV}$ in Ref.~\cite{Nieves:2012tt}. Therefore, we adopt the tripole form factor and set the range of $\alpha$ from $0.5$ to $1.5$ to obtain reasonable partial decay widths.

The decay width for the two-body processes $X_0 \to VV (PP)$ is expressed as
\begin{align}
    \Gamma= \frac{1}{\mathcal{S}}\frac{\vert \vec{p} \vert}{8 \pi m_{X_0}^2} \sum_{\rm spins} \left\vert \mathcal{M}_{V(P)}\right\vert^2,
\label{Eq.X2 decay rate}
\end{align}
where $\mathcal{S}$ is the symmetry factor. This factor is typically assigned a value of 1 unless the final state comprises identical particles, for which $\mathcal{S}$ is set to be 2. The symbol $\sum_{\mathrm{spin}}$ means the summation over the spins of final states. $\vec{p} $ denotes the three momentum of final state.

\section{Numerical Results And Discussions}\label{sec:results}

\begin{figure*}[htbp]
	\includegraphics[width=0.95\linewidth]{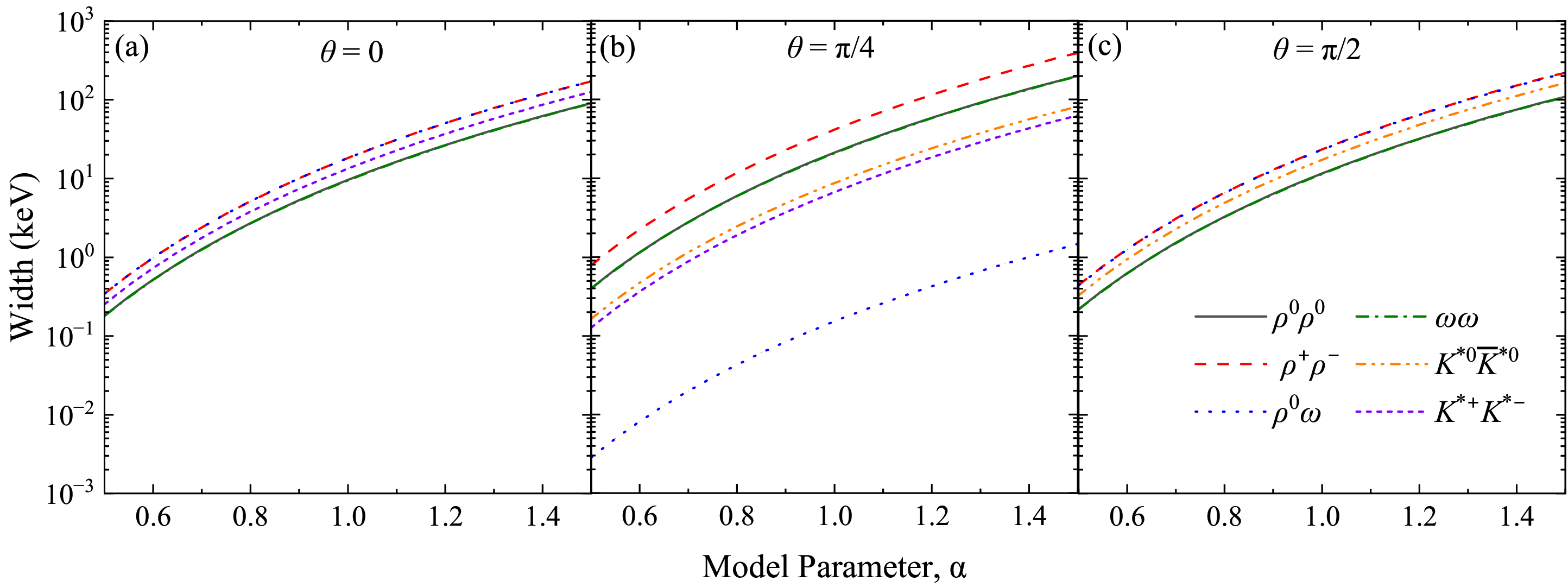}
	\caption{The dependence of the $X_0 \to V V$ decay widths on the model parameter $\alpha$ at three different phase angles $\theta = 0$ (a), $\pi/4$ (b), and $\pi/2$ (c).}
	\label{fig:VVθ03045alpha}
\end{figure*}

\begin{figure*}[htbp]
	\includegraphics[width=0.95\linewidth]{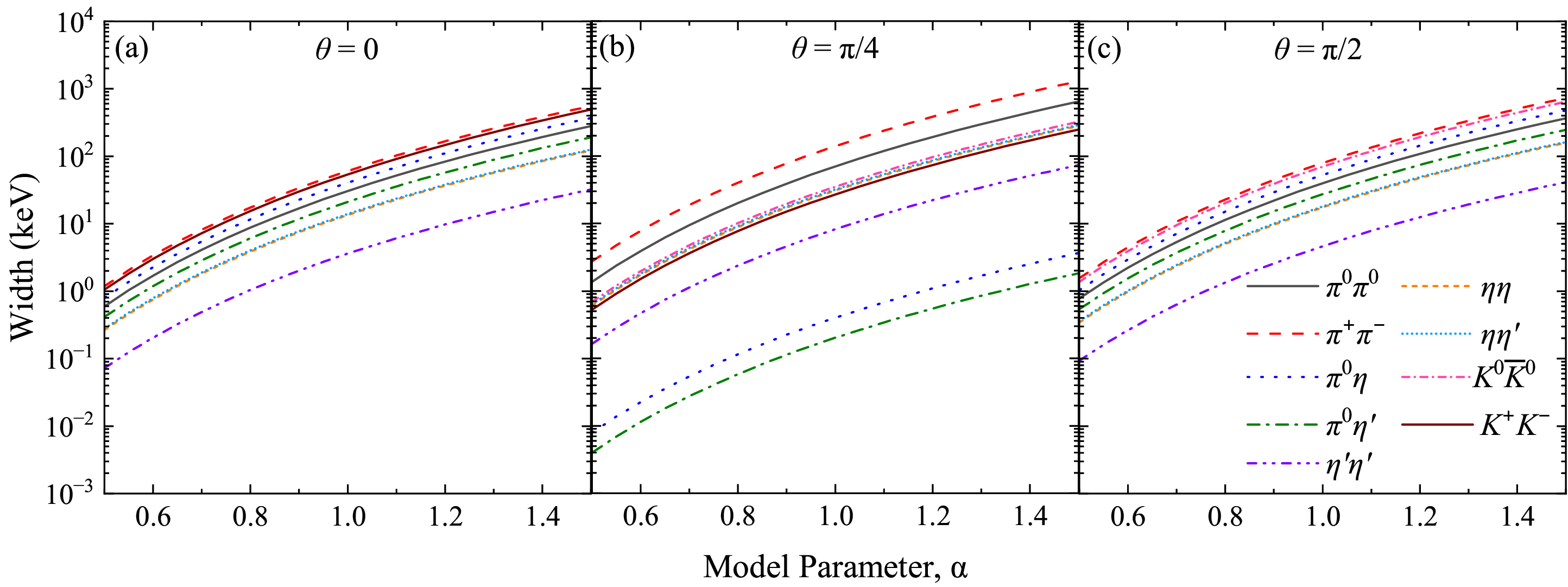}
	\caption{The dependence of the $X_0 \to P P$ decay widths on the model parameter $\alpha$ at three different phase angles $\theta = 0$ (a), $\pi/4$ (b), and $\pi/2$ (c). The $\eta$-$\eta^\prime$ mixing angle $\theta_\mathrm{P}=-19.1^\circ$ is taken from Refs.~\cite{MARK-III:1988crp,DM2:1988bfq}.} \label{fig:PPθ03045alpha}
\end{figure*} 

In this section, we first analyze the influence of different $\alpha$ values on the partial decay widths. The parameter $\alpha$ is varied from $0.5$ to $1.5$. We choose three different phase angles $\theta = 0$, $\pi/4$, and $\pi/2$. The $X_0$ is a pure neutral state when the phase angle $\theta = 0$. The $\theta = \pi/4$ describes a configuration where the neutral and charged components in the $X_0$ are of equal proportion. In the case of $\theta = \pi/2$, the $X_0$ only comprises the charged charmed mesons. 

Figure \ref{fig:VVθ03045alpha} shows the partial decay widths of the $X_0 \to V V$ as functions of the model parameter $\alpha$.
The decay of $X_0$ into the $K^{*0}\bar{K}^{*0}$ final state can occur only through the intermediate meson loop $[D^{+} D^{-}]D_s^{(*)+}$, so that the $X_0$ must contain charged component for this decay channel. Similarly, the decay into the $K^{*+}K^{*-}$ pair, involving the interaction of $[D^{0} \bar{D}^{0}]D_s^{(*)+}$, requires the $X_0$ to contain neutral component. This indicates that the $K^{*0}\bar{K}^{*0}$ decay channel is completely suppressed at the phase angle of $\theta=0$ [Fig.~\ref{fig:VVθ03045alpha} (a)], while the $K^{*+}K^{*-}$ channel is absent at $\theta=\pi/2$ [Fig.~\ref{fig:VVθ03045alpha} (c)]. Moreover, in these two cases, the isospin-violating effect originating from the difference between the $u$ and $d$ quark masses vanishes. Thus, the partial decay width of the $X_0 \to \rho^0 \omega$ nearly equals to that of the $X_0 \to \rho^+ \rho^-$. Because the couplings of the $X_0$ to the neutral and charged components are different [see Eqs. \eqref{eq:chi0} and \eqref{eq:chic}], the absolute widths for these two cases are different accordingly. At the phase angle of $\pi/4$, the neutral and charged components in the $X_0$ contribute equally. The decay width of the isospin-violating decay $X_0 \to \rho^0 \omega$ exhibits a dramatic suppression.

Overall, the partial decay widths for the $X_0\to VV$ increase as the model parameter $\alpha$ increases; the decays $X_0\to \rho^0\omega$, $K^{\ast 0}\bar{K}^{\ast 0}$, and $K^{\ast +}K^{\ast -}$ exhibit strong dependence on the phase angle $\theta$, while the other three decay channels $X_0\to \rho^0\rho^0$, $\rho^+\rho^-$, and $\omega\omega$ are weakly phase-angle-dependent. For the later channels, the predicted partial decay widths within the $\alpha$ range from $0.5$ to $1.5$ are
\begin{align}
    \Gamma(X_0\to \omega\omega)\approx \Gamma(X_0\to \rho^0\rho^0) &\approx \frac{1}{2} \Gamma(X_0\to \rho^+\rho^-)\nonumber\\
    &= (0.2 \sim 200 )~\mathrm{keV}.
\end{align}

In Fig. \ref{fig:PPθ03045alpha} we plot the partial decay widths of the $X_0\to PP$ as a function of the model parameter $\alpha$. As seen, the $PP$ channels exhibit similar behaviors with those of the $VV$ channels when the phase angle $\theta$ and model parameter $\alpha$ are varied. For the decay processes $X_0\to\pi^0\pi^0$, $\eta\eta$, $\eta\eta^\prime$, $\eta^\prime\eta^\prime$ and $\pi^+\pi^-$, which are nearly independent of the phase angle, the decay widths are 
\begin{subequations}
\begin{align}
    &\Gamma(X_0\to \pi^0\pi^0) = (0.6 \sim 600 )~\mathrm{keV}\, , \\
    &\Gamma(X_0\to \eta\eta) = (0.3 \sim 300 )~\mathrm{keV}\, , \\
    &\Gamma(X_0\to \eta\eta') = (0.3 \sim 300 )~\mathrm{keV}\, , \\
    &\Gamma(X_0\to \eta'\eta') = (0.07 \sim 80 )~\mathrm{keV}\, , \\
    &\Gamma(X_0\to \pi^+\pi^-) = (1 \sim 1300 )~\mathrm{keV}\, .
\end{align}
\end{subequations}

In Table \ref{tab:branchingratio} we list the partial decay widths for the concerned processes $X_0\to VV$ and $X_0\to PP$. In Ref. \cite{Gamermann:2009ouq} a hidden charm resonance with mass 3.722 GeV and width 36 MeV was predicted in a unitarized coupled channel framework. This dynamically generated scalar
resonance shows much stronger coupling to the $\eta^{(\prime)}\eta^{(\prime)}$ than to the $\pi\pi$, in contrast to our present results. The great difference might be useful for the future experiments, such as BESIII and Belle II, to distinguish the possible structures near 3.7 GeV.

\begin{table}[htbp]
	\caption{Decay widths (in units of keV) of the $X_0 \to V V$ and $P P$ for different phase angles $\theta =0$, $\pi/4$, and $\pi/2$. The $\eta$-$\eta^\prime$ mixing angle $\theta_\mathrm{P} = -19.1^\circ$ is taken from~\cite{MARK-III:1988crp,DM2:1988bfq} and the model parameter $\alpha$ ranges from 0.5 to 1.5.}
	\label{tab:branchingratio}
	\begin{ruledtabular}
		\begin{tabular}{llll}
		Final states	&$\theta = 0$&$\theta = \pi/4$&$\theta = \pi/2$ \\
			\colrule
			$\rho^0 \rho^0$ & $0.18-90.68$  & $0.40-199.96$ & $0.22-109.73$\\
			$\rho^+ \rho^-$ & $0.34-172.08$  & $0.78-392.39$ & $0.44-221.90$\\
			$\rho^0 \omega$ & $0.34-171.90$  & $0-1.46$ & $0.44-219.50$\\
			$\omega \omega$ & $0.18-89.67$  & $0.39-197.68$ & $0.22-108.46$\\
			$K^{*0} \bar{K}^{*0}$ & $\cdots$  & $0.16-81.87$ & $0.33-163.75$\\
			$K^{*+} K^{*-}$ & $0.25-125.91$  & $0.13-62.96$ & $\cdots$\\
                Total & $1.29-650.24$& $1.86-936.32$ &$1.65-823.34$\\ \colrule
			$\pi^0 \pi^0$ & $0.60-279.18$  & $1.37-639.98$ & $0.78-363.59$\\
			$\pi^+ \pi^-$ & $1.19-556.13$  & $2.73-1280.33$ & $1.56-730.11$ \\
                $\pi^0 \eta$ & $0.79-369.57$  & $0-3.64$  & 
            $1.03-480.51$\\
			$\pi^0 \eta'$ & $0.42-190.70$  & $0-1.82$ & $0.54-247.06$\\
			$\eta' \eta'$ & $0.07-32.15$  & $0.16-73.25$ & $0.09-41.39$ \\
                $\eta \eta$ & $0.26-122.18$  & $0.61-279.59$ & 
            $0.34-158.60$ \\	
			$\eta \eta'$ & $0.28-125.76$  & $0.63-287.19$ & $0.36-162.61$ \\
			$K^{0} \bar{K}^{0}$ & $\cdots$  & $0.69-320.72$ & $1.38-641.43$ \\
			$K^{+} K^{-}$ & $1.06-491.76$  & $0.53-245.88$ & $\cdots$\\
            Total & $4.67-2167.43$& $6.72-3132.40$ & $6.08-2825.3$\\
		\end{tabular}
	\end{ruledtabular}
\end{table}
 
\begin{figure*}[htbp]
	\centering
	\includegraphics[width=0.95\linewidth]{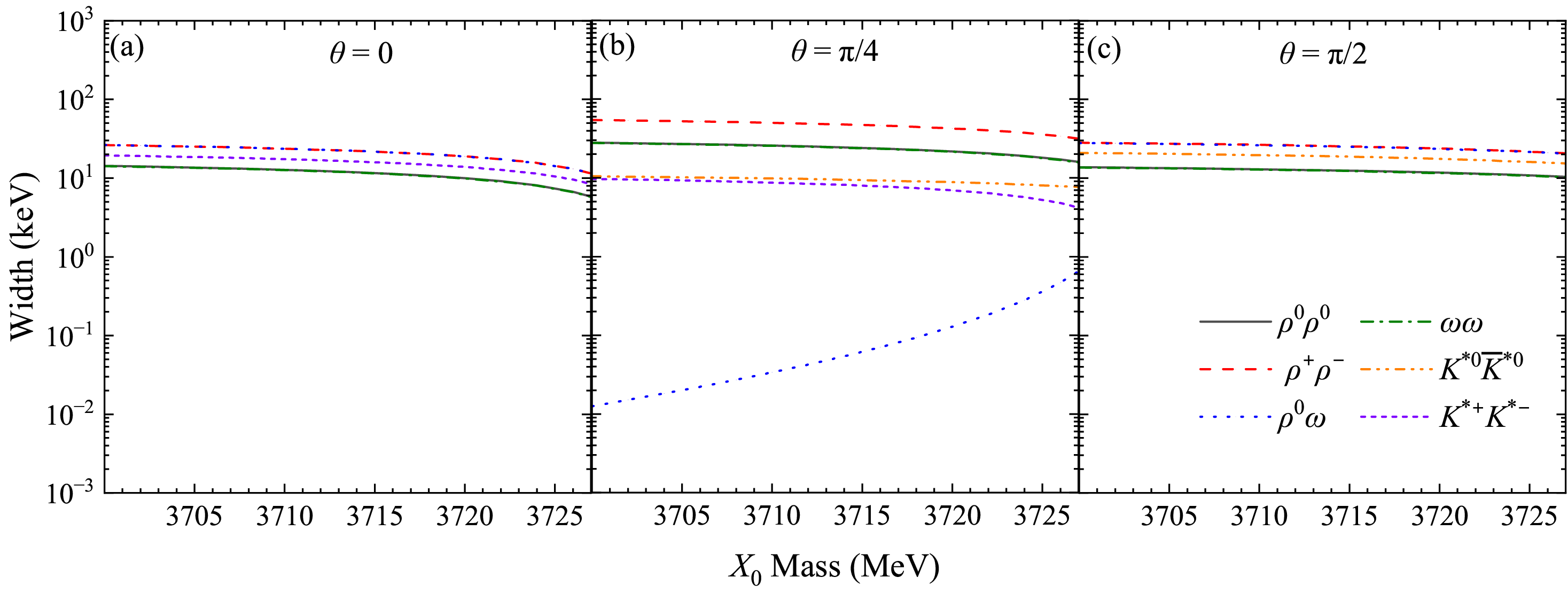}\,
	\caption{The $X_0$ mass dependence of the decay processes $X_0 \to V V$ for different phase angles $\theta=0$ (a), $\pi/4$ (b), and $\pi/2$ (c). The model parameter $\alpha$ is taken to be 1.0.}
	\label{fig:TMASSVV}
\end{figure*}

\begin{figure*}
	\centering
	\includegraphics[width=0.95\linewidth]{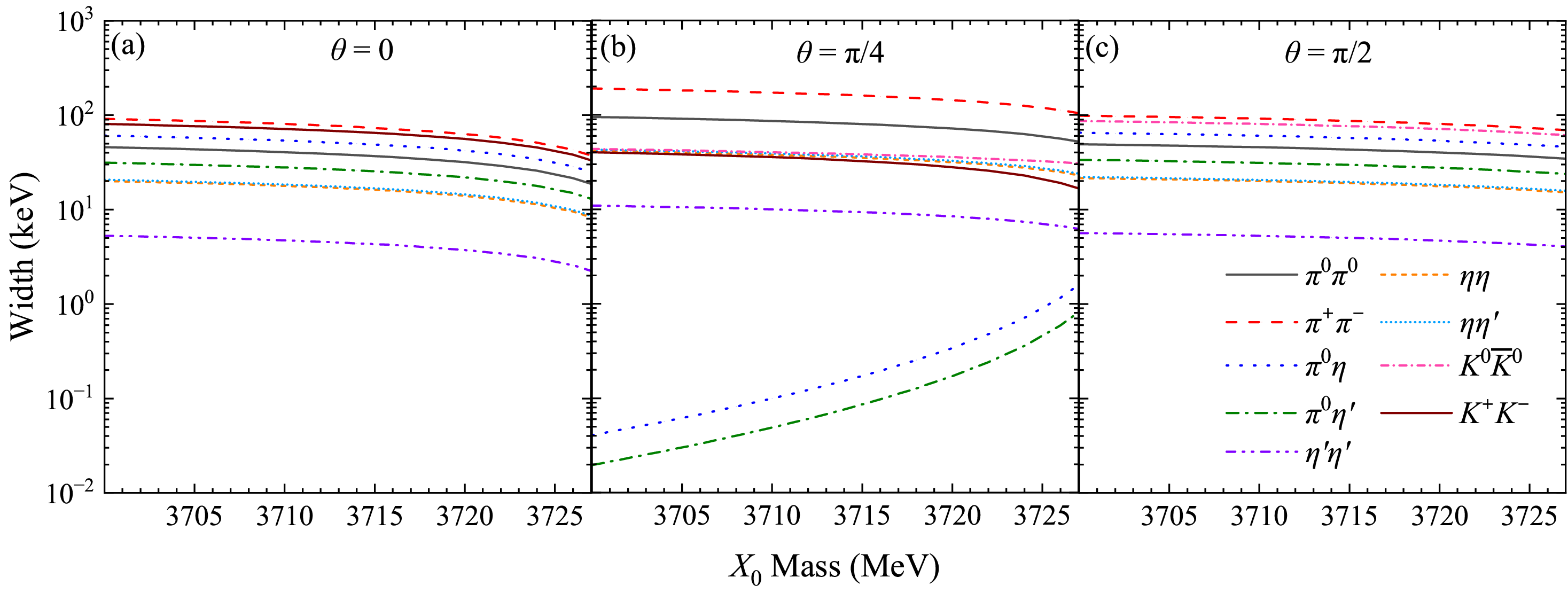}\,
	\caption{The $X_0$ mass dependence of the decay processes $X_0 \to PP$ for different phase angles $\theta=0$ (a), $\pi/4$ (b), and $\pi/2$ (c). The $\eta$-$\eta^\prime$ mixing angle $\theta_\mathrm{P} = -19.1^\circ$ is taken from Refs.~\cite{MARK-III:1988crp,DM2:1988bfq} and the model parameter $\alpha$ is taken to be 1.0.}
	\label{fig:PPm}
\end{figure*}

In order to investigate the influence of the $X_0$ mass on the decay widths, we varied the $X_0$ mass from $3700~\mathrm{MeV}$ to $3727~\mathrm{MeV}$~\cite{Hidalgo-Duque:2012rqv,Nieves:2012tt,Gamermann:2006nm,Deineka:2021aeu,Albaladejo:2013aka}. It is noted that this mass range is below the $D\bar{D}$ thresholds ($m_{D^0} + m_{{\bar D}^0} = 3729.68~\mathrm{MeV}$, $m_{D^+} + m_{D^-} = 3739.32~\mathrm{MeV}$). The calculated results for the partial decay widths of the $X_0\to VV$ and $PP$ at $\alpha=1.0$ are presented in Figs.~\ref{fig:TMASSVV} and \ref{fig:PPm}, respectively. It is seen that except the isospin-breaking decays $X_0 \to \rho^0\omega$ and $X_0\to \pi^0\eta(\eta^\prime)$, the widths of all the other decay processes exhibit slight decrease as the $X_0$ mass increases. This is because that the coupling strengths $\chi_{\mathrm{nr}}^0$ and $\chi_{\mathrm{nr}}^c$ governed by Eq.~\eqref{eq:chi} decrease with increasing the $X_0$ mass. However, the difference between the $\chi_{\mathrm{nr}}^0$ and $\chi_{\mathrm{nr}}^c$ becomes larger as the mass of $X_0$ grows. Therefore, the widths of the isospin-breaking decays $X_0 \to \rho^0\omega$ and $X_0\to \pi^0\eta(\eta^\prime)$ that are determined by the difference between neutral and charged meson loops increase clearly with increasing the $X_0$ mass.


From Figs.~\ref{fig:VVθ03045alpha} and \ref{fig:PPθ03045alpha} it is seen that the $\alpha$ dependence of the widths is similar for all cases. This model dependence could be cancelled for the width ratios. For the $X_0\to VV$ decay process, we define the following ratios:
\begin{subequations}\label{eq:ratio_VV}
\begin{align}
    R_1 &= \frac{\mathrm{\Gamma}(X_0 \to \rho^0\omega)}{\mathrm{\Gamma}(X_0 \to \omega\omega)} \, , \label{eq:ratio_VV1}\\
    R_2 &= \frac{\mathrm{\Gamma}(X_0 \to \rho\rho)}{\mathrm{\Gamma}(X_0 \to \omega\omega)}\, , \\
    R_3 &= \frac{\mathrm{\Gamma}(X_0 \to K^{*+}K^{*-})}{\mathrm{\Gamma}(X_0 \to \omega\omega)} \, , \\
    R_4 &= \frac{\mathrm{\Gamma}(X_0 \to K^{*0} {\bar K}^{*0})}{\mathrm{\Gamma}(X_0 \to \omega\omega)} \, ,
\end{align}
\end{subequations}
with $\Gamma(X_0\to \rho\rho)=\Gamma(X_0\to \rho^0\rho^0)+\Gamma(X_0\to \rho^+\rho^-)$.

Similarly, for $X_0 \to PP$, the $\pi^0\pi^0$ and $\pi^+\pi^-$ decay channels are classified into $\pi\pi$ and the relevant ratios are defined as follows:
\begin{subequations}\label{eq:ratio_PP}
\begin{align}
    r_1 &= \frac{\mathrm{\Gamma}(X_0 \to \pi^0 \eta)}{\mathrm{\Gamma}(X_0 \to \pi \pi)} \, , \\
    r_2 &= \frac{\mathrm{\Gamma}(X_0 \to \pi^0 \eta^\prime)}{\mathrm{\Gamma}(X_0 \to \pi \pi)} \, , \\
    r_3 &= \frac{\mathrm{\Gamma}(X_0 \to \eta \eta)}{\mathrm{\Gamma}(X_0 \to \pi \pi)} \, , \\
    r_4 &= \frac{\mathrm{\Gamma}(X_0 \to \eta \eta^\prime)}{\mathrm{\Gamma}(X_0 \to \pi \pi)} \, , \\
    r_5 &= \frac{\mathrm{\Gamma}(X_0 \to \eta^\prime \eta^\prime)}{\mathrm{\Gamma}(X_0 \to \pi \pi)} \, , \\
    r_6 &= \frac{\mathrm{\Gamma}(X_0 \to K^+ K^-)}{\mathrm{\Gamma}(X_0 \to \pi \pi)}\, ,\\
    r_7 &= \frac{\mathrm{\Gamma}(X_0 \to K^0 \bar{K}^0)}{\mathrm{\Gamma}(X_0 \to \pi \pi)}\,.
\end{align}
\end{subequations}

\begin{figure}
	\centering
	\includegraphics[width=0.94\linewidth]{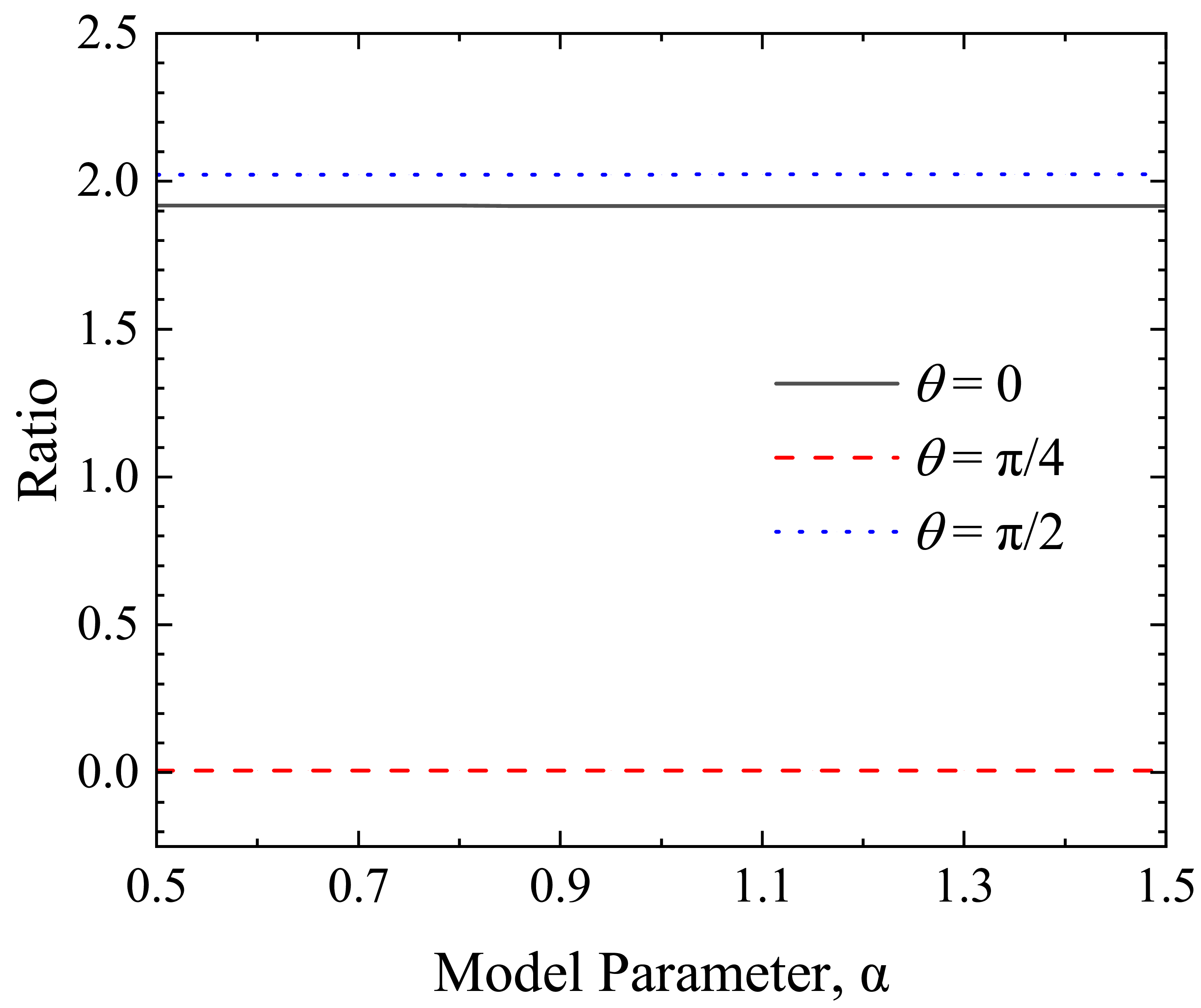}
	\caption{The $\alpha$ dependence of the ratio $R_1$ defined in Eq.~(\ref{eq:ratio_VV1}).}
	\label{fig:VV_PPalpha}
\end{figure}

\begin{figure}
	\centering
	\includegraphics[width=0.94\linewidth]{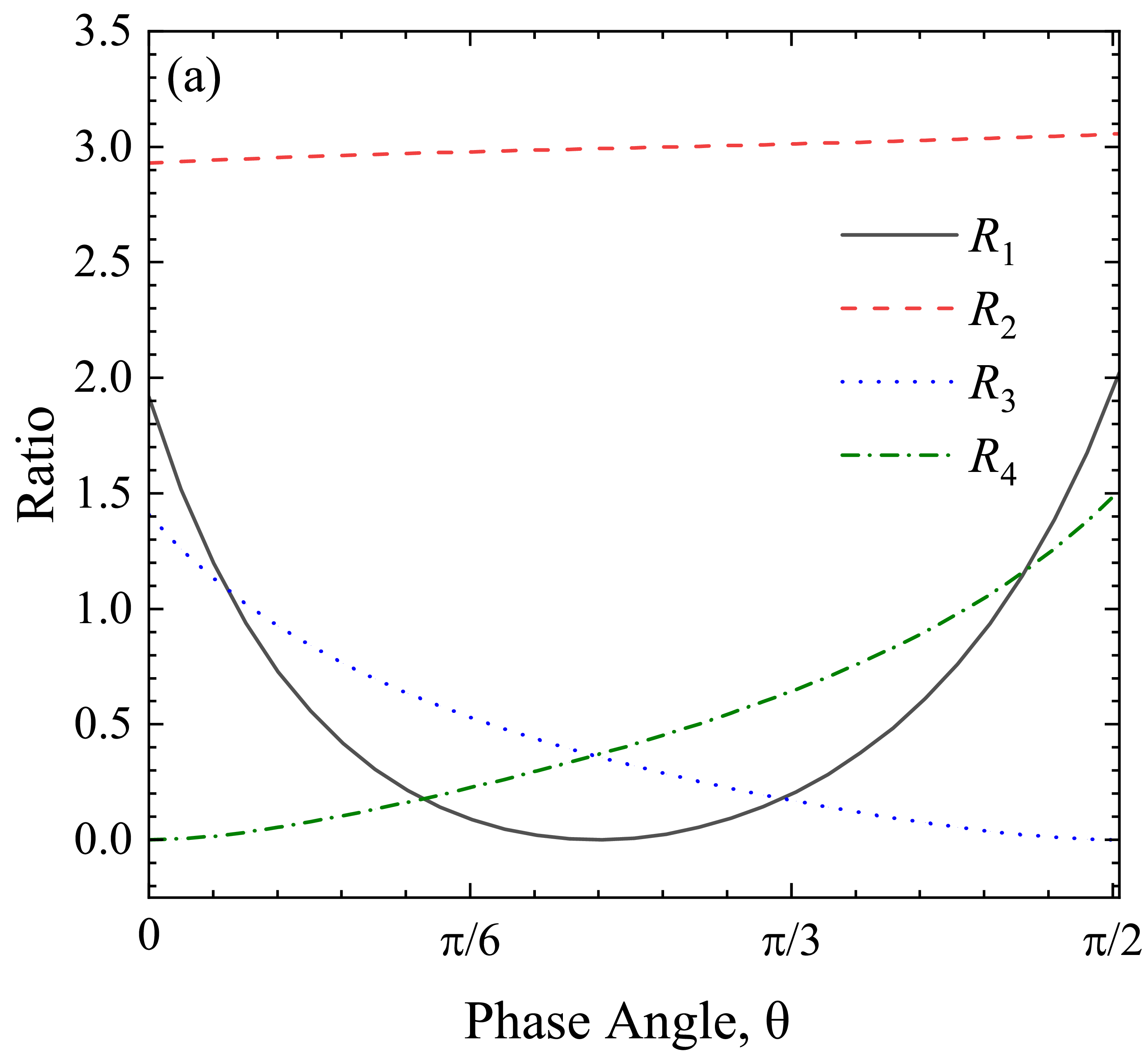}
	\includegraphics[width=0.94\linewidth]{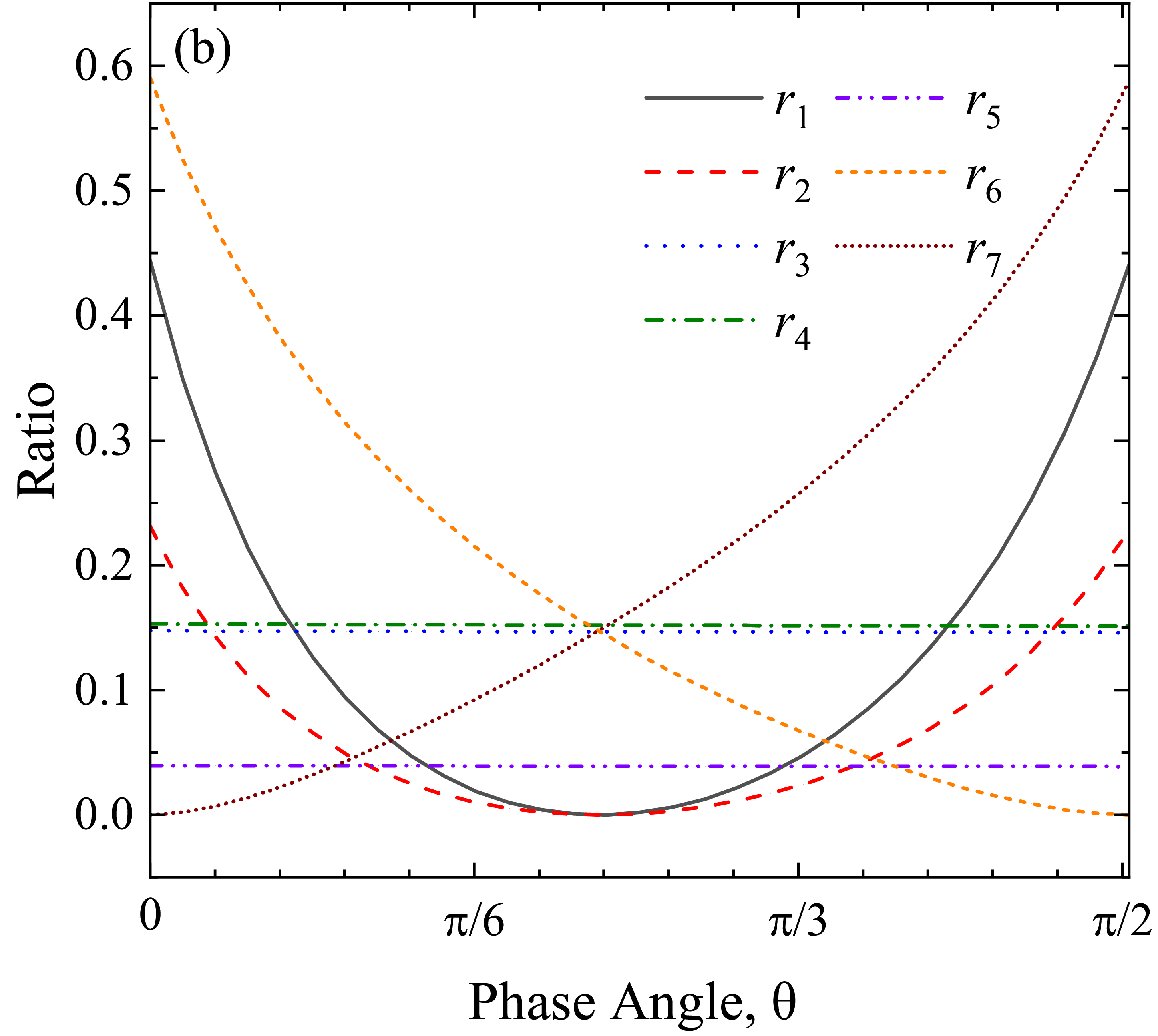}\,
	\caption{Ratios $R_i$ (a) defined in Eq.~(\ref{eq:ratio_VV}) and $r_i$ (b) defined in Eq.~(\ref{eq:ratio_PP}) as a function of the phase angle $\theta$. The $\eta$-$\eta^\prime$ mixing angle $\theta_\mathrm{P} = -19.1^\circ$ is taken from Refs.~\cite{MARK-III:1988crp, DM2:1988bfq} and the model parameter $\alpha=1.0$.}
	\label{fig:VV_PPtheta}
\end{figure}

The dependence of the ratio $R_1$ on the model parameter $\alpha$ is shown in Fig.~\ref{fig:VV_PPalpha}. It is evident that $R_1$ is rather insensitive to the $\alpha$. We have also investigated the dependence of the ratio $R_1$ on the model parameter $\alpha$ using monopole and dipole form factors, finding that $R_1$ remains nearly independent of the model parameter $\alpha$. Moreover, the ratios are approximately equal to those obtained using the tripole form factor. The consistency of the ratio with respect to $\alpha$ suggests that the form factor effectively regulates the cutoff for each channel to a certain extent. However, the ratio exhibits discrepancies under different phase angles, which prompts us to continue the discussion on the phase-angle dependence.

Since the phase angle characterizes the molecular configuration of the $X_0$, we investigate the ratios $R_i$ and $r_i$ for different phase angles. In Fig.~\ref{fig:VV_PPtheta}, we plot the ratios $R_i$ defined in Eq.~(\ref{eq:ratio_VV}) and $r_i$ in Eq.~(\ref{eq:ratio_PP}) as a function of the phase angle $\theta$. The results were obtained using $\alpha=1.0$. The trends in the ratios can be broadly categorized into the following four cases. The ratios $R_2$ and $r_{3,4,5}$ are nearly independent of the phase angle $\theta$. The ratios $R_3$ and $r_6$ decrease with increasing $\theta$, while the ratios $R_4$ and $r_7$ increase  monotonically as the phase angle $\theta$ grows. Within the selected phase angle range, the ratios $R_1$ and $r_{1,2}$ exhibit a non-monotonic behavior characterized by an initial decrease followed by a subsequent increase. These ratios shown in Fig.~\ref{fig:VV_PPtheta} may be tested by the future experimental measurements at BESIII and Belle II and can be used to determine the value of the phase angle.

\section{Summary}\label{sec:summary}

In this work, we investigated the hadronic decays of the spin-0 partner of $X(3872)$ (called $X_0$) into two light hadrons. According to the assumption of the $X_0$ as a $D\bar{D}$ molecular state, we calculated the partial widths of the $X_0 \to V V$ and $P P$ decay processes using an effective Lagrangian approach. We considered three cases: pure neutral state ($\theta = 0$), isospin singlet state ($\theta = \pi/4$) and pure charged state ($\theta = \pi/2$), where the phase angle  $\theta$ determines the proportion of the neutral and charged components in the $X_0$. 
The decay widths depend on the model parameter $\alpha$. However, the relative width ratios are found to be nearly model-$\alpha$-independent. As the $\alpha$ increases from $0.5$ to $1.5$, the total width of the $X_0 \to VV$ are between a few $\mathrm{keV}$ and hundreds of $\mathrm{keV}$, while the total width of the $X_0 \to PP$ varies from a few $\mathrm{keV}$ to several $\mathrm{MeV}$. 

Moreover, we also investigate the influence of the $X_0$ mass on the decay widths. Based on previous theoretical predictions, we vary the $X_0$ mass from $3700~\mathrm{MeV}$ to $3727~\mathrm{MeV}$, which are bellow the $D\bar{D}$ threshold. It was found that the widths for the isospin-breaking decays $X_0 \to \rho^0\omega$ and $X_0\to \pi^0\eta(\eta^\prime)$ increase clearly as the $X_0$ mass goes up, while the widths for the other decays decrease slightly. Finally, the relative width ratios between different channels are studied to reflect the influence of the molecular configuration of the $X_0$. We hope that our calculated results are helpful for searching the $X_0$ in the future experiments at BESIII and Belle II.

\begin{acknowledgements}\label{sec:acknowledgements}

This work is partly supported by the National Natural Science Foundation of China under Grant Nos. 12475081, 12105153, and by the Natural Science Foundation of Shandong Province under Grant Nos. ZR2021MA082, and ZR2022ZD26. It is also supported by Taishan Scholar Project of Shandong Province (Grant No. tsqn202103062).

\end{acknowledgements}

\bibliography{X0.bib}
\end{document}